\documentclass[sigconf,natbib=true,anonymous=false]{acmart}

\usepackage{amssymb}
\usepackage{array}
\usepackage{url}
\usepackage{dblfloatfix}
\usepackage{booktabs}
\usepackage{amsmath}
\usepackage{tabularx}
\usepackage{colortbl}

\usepackage{graphics}
\usepackage{epsfig}
\usepackage{multirow}
\usepackage{flushend}

\usepackage{url}
\usepackage{tikz}
\usepackage{enumitem}
\usepackage{multicol}

\usepackage{verbatim}
\usepackage{float}
\usepackage{subcaption}

\usepackage{algorithm}
\usepackage{algpseudocode}

\usepackage{marginnote}
\usepackage{xspace}
\usepackage{balance}

\usepackage{todonotes}
\usepackage{xcolor}

\newcommand{\squishlist}{
 \begin{list}{$\bullet$}
  { \setlength{\itemsep}{0pt}
     \setlength{\parsep}{1pt}
     \setlength{\topsep}{1pt}
     \setlength{\partopsep}{0pt}
     \setlength{\leftmargin}{1.5em}
     \setlength{\labelwidth}{1em}
     \setlength{\labelsep}{0.5em} } }

\newcommand{\squishend}{
  \end{list}  }

\author{Iain Mackie}
\affiliation{
  \institution{University of Glasgow}
}
\email{i.mackie.1@research.gla.ac.uk}

\author{Shubham Chatterjee}
\affiliation{
  \institution{University of Glasgow}
}
\email{shubham.chatterjee@glasgow.ac.uk	}

\author{Sean MacAvaney}
\affiliation{
  \institution{University of Glasgow}
}
\email{sean.macavaney@glasgow.ac.uk}

\author{Jeffrey Dalton}
\affiliation{
  \institution{University of Glasgow}
}
\email{jeff.dalton@glasgow.ac.uk}

\renewcommand\footnotetextcopyrightpermission[1]{} 

\begin{document}
\fancyhead{}

\title{Adaptive Latent Entity Expansion for Document Retrieval}

\begin{abstract}

Despite considerable progress in neural relevance ranking techniques, search engines still struggle to process complex queries effectively --- both in terms of precision and recall. Sparse and dense Pseudo-Relevance Feedback (PRF) approaches have the potential to overcome limitations in recall, but are only effective with high precision in the top ranks. 
In this work, we tackle the problem of search over complex queries using three complementary techniques. First, we demonstrate that applying a strong neural re-ranker before sparse or dense PRF can improve the retrieval effectiveness by 5--8\%. This improvement in PRF effectiveness can be attributed directly to improving the precision of the feedback set. Second, we propose an enhanced expansion model, Latent Entity Expansion (LEE), which applies fine-grained word and entity-based relevance modelling incorporating localized features. Specifically, we find that by including both words and entities for expansion achieve a further 2--8\% improvement in NDCG. Our analysis also demonstrated that LEE is largely robust to its parameters across datasets and performs well on entity-centric queries. And third, we include an ``adaptive'' component in the retrieval process, which iteratively refines the re-ranking pool during scoring using the expansion model and avoids re-ranking additional documents. We find that this combination of techniques achieves the best NDCG, MAP and R@1000 results on the TREC Robust 2004 and CODEC document datasets, demonstrating a significant advancement in expansion effectiveness.

\end{abstract}

\maketitle

\section{Introduction}
\label{sec:intro}

A fundamental problem in information retrieval is query-document lexical mismatch~\cite{belkin1982ask}.  
A common approach to address this issue is Pseudo-Relevance Feedback (PRF), where a first-pass top-$k$ candidate set of documents is retrieved, and these feedback signals can augment the query for a second-pass retrieval.
Early work on PRF focused on query expansion, including Relevance Modeling~\cite{metzler2005markov, abdul2004umass} as well as extensions like Latent Concept Expansion~\cite{metzler2007latent} and Collection Enrichment~\cite{Kwok1998ImprovingTA}. A significant advance was the use of entity-based representations, demonstrating improvements over term-based models~\cite{dalton2014entity}.
Recently, this PRF paradigm has also leveraged dense vectors ~\cite{naseri2021ceqe, wang2022colbert, yu2021improving}. 
However, all these models suffer from the same problem: If the initial query is challenging, the candidate set is unlikely to contain relevant documents in the top ranks, which will cause PRF models to fail.

\begin{figure}[h!]
    \centering
    
    \setlength{\belowcaptionskip}{-5pt}
    \includegraphics[scale=0.1]{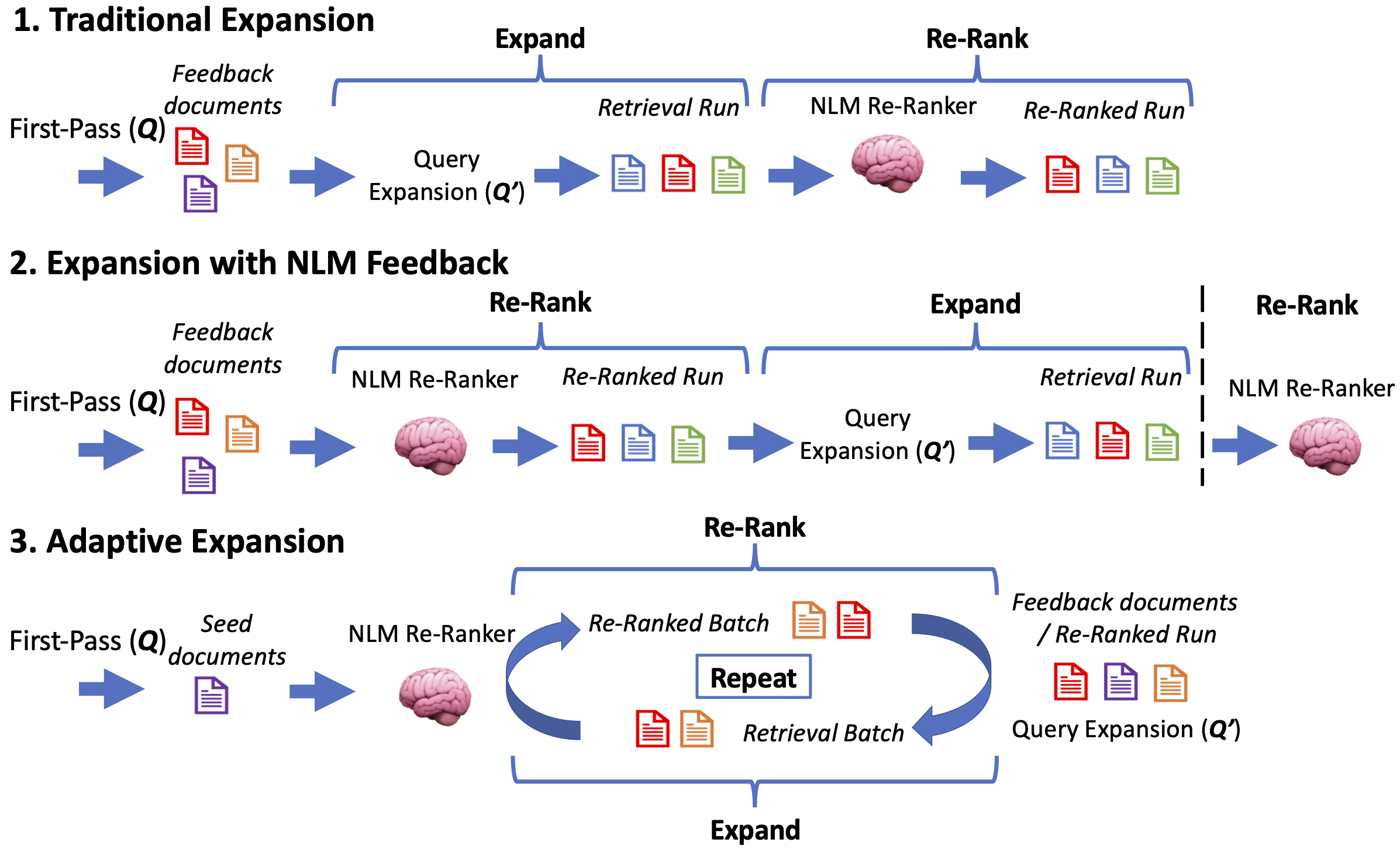}
    \vspace{-2em}
    \caption{
     Adaptive expansion: \textit{1) Traditional Expansion:} standard first-pass and expand approach, often with neural re-ranking after. \textit{2) Expansion with NLM Feedback:} a re-ranked run provides more accurate feedback for expansion. \textit{3) Adaptive Expansion:}  iteratively re-rank batches of documents before issuing expanded query to retrieve the next batch.    
    }
    \label{img:rre}
\end{figure}

Meanwhile, neural language models (NLMs) for re-ranking~\cite{lin2022pretrained} have led to significant advances in effectiveness, particularly precision in the top ranks. 
In this work, we pull together these research threads on neural re-ranking and entity-based expansion methods to improve the core task of document retrieval. 
Figure~\ref{img:rre} shows how we address the problem of poor pseudo-relevance feedback by applying re-ranking prior to query expansion and re-executing this query. 
We find that expansion with NLM feedback improves the recall-oriented effectiveness of sparse and dense PRF approaches. 

Armed with insights from this analysis, we propose a new model to improve PRF effectiveness further when operating over NLM feedback: Latent Entity Expansion (LEE). LEE is a joint probabilistic term and entity-based expansion model. In contrast with prior work in Latent Concept Expansion, we show that a hybrid expansion model with terms and entities is more effective than comparable individual expansion models. We also demonstrate improved effectiveness from passages based on NLM re-ranking that provide a more fine-grained hybrid relevance model. 
Furthermore, unlike prior work~\cite{metzler2007latent}, we find that using dependencies from entity co-occurrence improves effectiveness with passage feedback, but can be harmful with document feedback.

Nonetheless, after our NLM expansion, we find that a second round of neural re-ranking is required to maximize precision. 
Thus, we draw inspiration from recent adaptive re-ranking work~\cite{sean2022adaptive} and propose our proposed ``adaptive expansion'' framework.  
Specifically, Figure~\ref{img:rre} shows how we dynamically refine the re-ranking pool during scoring using the expansion model.
This allows us to use NLM feedback for expansion and re-ranking in a single pass and reduces the number of documents scored by around 35\%.

Our document test collections, namely TREC Robust and CODEC, focus on challenging ``complex'' queries. Unlike recent web collections that emphasize ``easy'' factoid-focused queries, these collections represent challenging topics where existing state-of-the-art methods for sparse and dense retrieval still have significant headroom~\cite{mackie2021deep, bolotova2022non}. 
By ``complex'' queries, we refer to queries which require significant amounts of context and reasoning capabilities~\cite{mackie2022codec}, as well as an understanding of multiple concepts, for example, ``Lasting social changes brought about by the Black Death''. 

Through extensive experiments under various conditions, LEE significantly improves core document retrieval effectiveness over previous expansion approaches. To our knowledge, it produces the highest recall ever achieved on these benchmark datasets by 6-12\%. 
Query analysis shows that LEE's hybrid expansion model with terms and entities improves the hardest entity-centric queries, where a fine-grained relevance model and entity dependencies are particularly useful.
Furthermore, LEE with adaptive expansion sets a new state-of-the-art for MAP and NDCG without requiring a second round of neural re-ranking, and our model parameters are robust across datasets. 
Overall, this work demonstrates the potential of probabilistic term-entity expansion models when combined with neural re-ranking.
We summarize our contributions below:

\begin{itemize}[leftmargin=*] 

\item We provide a detailed study of existing probabilistic word and entity expansion models when combined with NLM re-rankers with document and passages feedback.

\item We propose a new hybrid relevance model for query expansion that incorporates entity dependencies.

\item We show that our unsupervised expansion model is state-of-the-art by 6-12\% on recall, and when combined with additional neural re-ranking, result in 2-8\% improvement on NDCG and MAP. 

\item We show that our hybrid relevance model with adaptive expansion achieves similar effectiveness improvements without additional NLM re-ranking (saving around 35\% compute).

\end{itemize}

\section{Related Work}
\label{sec:Related Work}


\subsection{Query Expansion}

Query expansion~\cite{rocchio1971relevance} deals with lexical mismatch~\cite{belkin1982ask} by expanding the query with terms closer to the underlying information need. A common automatic approach is \textit{pseudo-relevance feedback} where the top-$k$ documents from an initial retrieval set are assumed relevant. Famous classical methods include Rocchio~\cite{rocchio1971relevance}, KL expansion~\cite{zhai2001model}, Relevance Modelling \cite{metzler2005markov}, and RM3 expansion~\cite{abdul2004umass}.
Furthermore, recent work, such as CEQE~\cite{naseri2021ceqe}, uses query-focused contextualized embedding for expansion. 
Conversely, this work evaluates expansion models based on NLM re-ranked feedback.
Specifically, LEE builds upon Latent Concept Expansion (LCE)~\cite{metzler2007latent} to develop a hybrid probability distribution over both words and entities based on re-ranked passage feedback, incorporating entity dependencies. 

The rise of dense retrieval has brought variants using vector-based PRF models~\cite{li2022improving}, including ColBERT PRF~\cite{wang2022colbert}, and ColBERT-TCT PRF~\cite{lin-etal-2021-batch}, and ANCE PRF~\cite{yu2021improving}. \citet{yu2021pgt} also shows that Transformer-based re-rankers can benefit from the extra context.
However, our results shows sparse and entity-based approaches are currently more effective than dense models for document retrieval on complex topics.
Nonetheless, we do find that using NLM feedback for dense retrieval improves recall.  

Recent work shows that ``adaptive re-ranking'' improves the effectiveness of passage ranking with minimal drawbacks on efficiency~\cite{sean2022adaptive}. Specifically, instead of independent first-pass and second-pass retrieval, \citet{sean2022adaptive} updates the re-ranking pool after each batch of NLM re-ranking using the Cluster Hypothesis~\cite{jardine1971use} to prioritise documents similar to those highly scored. Drawing inspiration from this research direction, we introduce ``adaptive expansion'' and show that iteratively updating our LEE query after each batch is significantly more effective than prior adaptive re-ranking methods for document ranking. 


\subsection{Entity-Centric Ranking}
\label{subsec:Entity-Centric Query Expansion}

Our work builds on and extends extensive research that incorporates entity-based representations within document ranking~\cite{shehata2022early, meij2010conceptual, xiong2015query, shehata2022early, xiong2016bag,  liu2015latent, xiong2015esdrank, liu2014exploiting}.
Prior work typically uses entity mentions present in the query or documents to help ground the task to an external Knowledge Base (KB), where entity linking is the process of identifying entity mentions within documents \cite{ferragina2011fast, piccinno2014tagme, li2020efficient, van2020rel}. 

Prior work has used entity-based query expansion methods to enrich the query with useful concepts to help retrieve relevant documents~\cite{meij2010conceptual, xiong2015query, Xu2009query}.
Furthermore, \citet{raviv2016document} developed an entity-based language model for document retrieval by representing queries and documents as a bag-of-words and bag-of-entity-links.
EQFE~\cite{dalton2014entity} enriches the query with KG entity-based features to improve document ranking, improving the hardest topics. 
Moreover, the Word-Entity Duet~\cite{xiong2017word} framework uses word-based and entity-based representations to embed documents and queries for ad-hoc retrieval.
Lastly, recent work of~\citet{shehata2022early} shows that enriching queries and documents using a dense end-to-end entity linking system~\cite{li2020efficient} can provide knowledge-grounded context and improve initial retrieval.
Our work builds on this literature by proposing a hybrid word and entity relevance model derived using NLM passage feedback.
Furthermore, we incorporate localization features such as entity dependence.
We show that entities are beneficial when used with words (and actually competitive by themselves) given our strong NLM passage ranking, significantly outperforming all prior methods.

\subsection{NLM Document Ranking}


The emergence of neural language model (NLM)~\cite{lin2022pretrained} has shown improvement across information retrieval tasks.
However, NLM re-rankers cannot easily ingest the full text when ranking long documents due to the input constraints of these models.
Various strategies deal with this problem; for example, BERT-maxp~\cite{dai2019deeper, yilmaz2019applying} and T5-maxp~\cite{nogueira2020document} shard long documents into passages that the model can score individually, and the highest-scoring passage is used as a proxy for overall document relevance.
In our work, we build upon similar intuition and use NLM passages ranking~\cite{nogueira2020document} to identify the most relevant sections of documents to form a more fine-grained relevance model.
Additionally, the precision improvements in the top ranks due to NLMs~\cite{craswell2020overview, craswell2021overview} provides a more accurate PRF for our expansion models.

We also compare our results against state-of-the-art models fine-tuned on the target datasets.  
For example, CEDR-KNRM~\cite{macavaney2019cedr} proposes a joint approach combining BERT’s passage outputs and handling passage aggregation via a representation aggregation technique.
Meanwhile, PARADE~\cite{li2020parade} uses attention to aggregate passage-level relevance signals using ELECTRA-based~\cite{clark2020electra} embeddings.
Lastly, MORES~\cite{gao2022long} is a modular re-ranker framework that uses  BART~\cite{lewis2020bart} for query-to-document cross attention.

\section{Adaptive Expansion}
\label{sec:method}





\begin{figure*}[h!]
    \centering
    \setlength{\abovecaptionskip}{-0.5pt}
    \setlength{\belowcaptionskip}{-5pt}
    \includegraphics[scale=0.14]{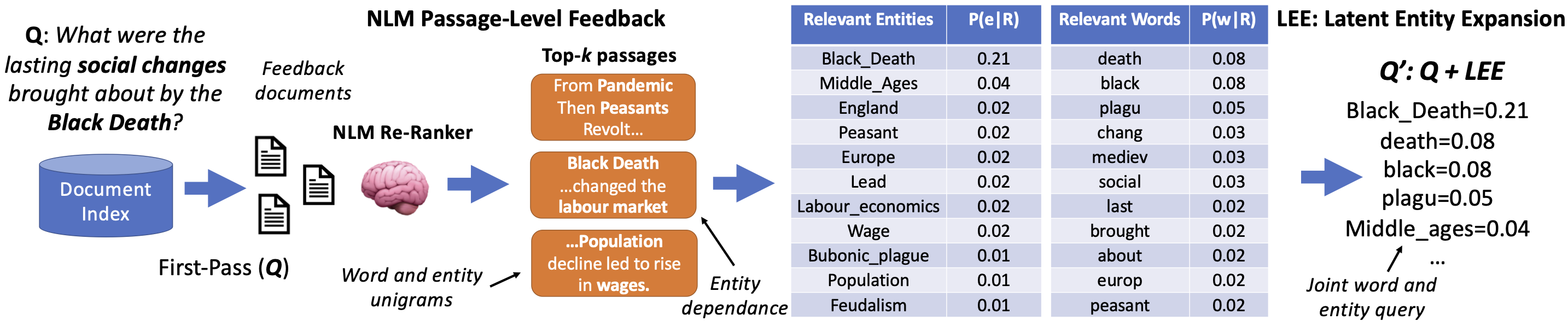}
    \caption{Overview of LEE:
    A NLM re-ranker supplies a strong passage ranking to provide fine-grained feedback.   
    LEE uses word and entity unigrams and entity dependencies to construct a hybrid word and entity-based probability distribution. 
    }
    \label{img:lee-overview}
\end{figure*}

\subsection{Rethinking Expansion Pipelines}
\label{sec:pipeline}

Based on the analysis of current retrieval models, we rethink the standard query expansion pipeline drawing on several research threads. 
Specifically, NLM re-ranking models~\cite{dai2019deeper, nogueira2020document, nogueira2019passage} offer an opportunity to improve the precision of document feedback to form more effective expansion models.
We also draw from recent work on adaptive re-ranking~\cite{sean2022adaptive} to allow our expansion model to use NLM feedback without incurring additional re-ranking cost. 

Formally, given a information need (query) $Q$, we want to return a ranked list of documents $D = [D_1, D_2,...,D_N]$ relevant to the query $Q$ from a collection $C$. For generality, documents, $D$, may also refer to other retrieval units, such as passages. 
We abstract a document ranking pipeline, and focus on changing the ordering of \textit{query expansion} and \textit{neural re-ranking} components. Figure \ref{img:rre} shows the three expansion pipelines we explore:  

\begin{itemize}[leftmargin=*] 

\item \textbf{Traditional expansion}: This is our standard document ranking pipeline with expansion~\cite{li2020parade, macavaney2019cedr, gao2022long}. Specifically, we retrieve an initial feedback set of documents from the collection before expanding the query to retrieve a new list of documents. Documents are then re-scored using a neural re-ranker to create our final re-ranked list of documents. The issue with this pipeline is that signals from advanced neural re-rankers~\cite{li2020parade, macavaney2019cedr, nogueira2020document} are not used to improve initial recall.

\item \textbf{Expansion with NLM feedback}: This is where we move NLM re-ranking before our expansion model in the pipeline to improve the precision of the feedback set; thus, improving expansion effectiveness. Additionally, we could further improve the precision of the final document ranking via a second re-ranking pass; however, this would also increase computational expense due to extra document scoring.  

\item \textbf{Adaptive expansion}: Instead of having a static run that we re-rank, we propose dynamically updating our document frontier as more documents are scored using our query expansion model. Specifically, we alternate our re-ranking of documents between the initial retrieval documents and the dynamic frontier based on the expansion model, which uses all currently re-ranked documents as its feedback set. This iterative batch process of re-ranking and expansion continues, with a batch size of $b$, until we reach our intended number of documents. Intuitively, updating our query expansion model as more documents are scored is similar to a manual researcher building their understanding of a topic through reading information. Additionally, unlike expansion with NLM feedback with a second re-ranking pass, adaptive expansion does not require additional computation from document re-ranking.

\end{itemize}

\subsection{Latent Entity Expansion}
\label{subsec:Deriving Expansion Words}

Figure~\ref{img:lee-overview} depicts our Latent Entity Expansion model that incorporates words and entities. 
Specifically, our query expansion approach uses a strong NLM re-ranked list of documents, which benefits precision in the top ranks (making our feedback more accurate). 
Thus, we assume top-$k$ documents to be query-relevant feedback $R$, which we use to construct Latent Entity Expansion (LEE) based on a hybrid relevance model of words ($\{w_1,w_2,...,w_i\} \in D$) and entities ($\{e_1,e_2,...,e_N\} \in D$).
We use LEE to expand the initial, $Q \to Q_{LEE}$, and retrieve our final set of documents, $[D_1, D_2,...,D_N]$.

\subsubsection{Deriving Expansion Words}
\label{subsec:Deriving Expansion Words}

Equation~\ref{eq:unigram-equation} show how we estimate the probability of a word $P(w|R)$ given the assumed relevant documents $R$. 
$P(Q|D)$ is obtained by normalizing the NLM re-ranking scores~\cite{nogueira2020document}, before we turn it into a probability by dividing the sum of all the normalized scores, $\sum_{D' \in R} P(Q|D')$. 
The probability of a word given a document, $P(w|D)$, is the term frequency divided by the document length. 
Following LCE~\cite{metzler2007latent}, we normalize the distribution using $P(w|C)$ (that we approximate for convenience with  $\text{IDF}(w,C)$). Later, in Section \ref{sec:RQ2}, we show that this feature is important for modelling the relevance of documents.

\begin{equation}
  \centering
   \label{eq:unigram-equation}
  P(w|R) = \sum_{D \in R} \frac{P(Q|D)}{\sum_{D' \in R} P(Q|D')} P(w|D) P(w|C)
\end{equation}

\subsubsection{Deriving Expansion Entities}
\label{subsec:Deriving Expansion Entities}

Analogously, we estimate the query-relevance of a document based on the entities contained within that document ($e \in D$). Prior work~\cite{metzler2007latent} only uses unigram representations because word dependencies do not improve results. In contrast, LEE incorporates both entity unigrams and dependencies and finds meaningful improvement in effectiveness with passage NLM feedback. 
The base formulation for entity terms follows how we model word unigrams, providing a unigram estimate of $P(e|R)$. However, we also include entity dependence terms based on co-occurrence to model the relationship between entities. 

\textbf{Estimating relevance of entity dependence}, we estimate this probability as follows: 

\begin{equation}
  \centering
   \label{eq:co-occur-equation}
  P([e_1, e_2]|R) = \sum_{D \in R} \frac{P(Q|D)}{\sum_{D' \in R} P(Q|D')} P([e_1, e_2]|D)  P([e_1, e_2]|C)
\end{equation}
where $P(Q|D)$ is the normalised NLM score and $P([e_1, e_2]|D)$ is the sum of both entity frequencies divided by the document length. We approximate $P([e_1, e_2]|C)$ as the product of entity IDFs, $\text{IDF}(e_1) \cdot \text{IDF}(e_2)$. Incorporating entity co-occurrence increases the weighting of entities that co-occur with many entities in relevant documents. This helps prioritise the ``central entities'' that are particularly useful for identifying relevant documents. Unlike \cite{metzler2007latent}, results show this entity dependence feature is particularly beneficial with passage feedback, although not meaningful at a document level.

We then combine the unigram and entity dependence models as follows. 
Mathematically, we model $P(e|R)$ as follows: 
\begin{equation}
  \centering
   \label{eq:entity-equation}
  P(e|R) =  \beta \sum_{e_i \in R} P([e, e_i]|R) + (1- \beta) P_{\text{unigram}}(e|R)
\end{equation}
where $P([e, e_i]|R)$ is the probability of the entity pair $(e, e_i)$ being in a relevant document, and $P_{\text{unigram}}(e|R)$ is probability of entity $e$, obtained using a unigram language model. 

\subsubsection{LEE Duet Representation}
\label{subsec:Document Ranking using LEE}

The final score of a document $D \in R$ is derived from an interpolation of the term-based and entity-based query expansion retrieval scores:
\begin{equation}
    \label{eq:final-score}
    \text{Score}(D, Q) = \lambda \cdot \text{Score}_\text{word}(D,Q) + (1 - \lambda) \cdot \text{Score}_\text{entity}(D,Q)
\end{equation}
where $\text{Score}_\text{word}(D,Q)$ is the document score based on our word query expansion, $\text{Score}_\text{entity}(D,Q)$ is the document score based on our entity query expansion. 
For simplicity to execute over large collections, we use BM25~\cite{robertson1994some} to execute our probabilistic queries over separate document and entity indexes.
Furthermore, following work by RM3~\cite{abdul2004umass}, we also include the probabilistic interpolations between the terms in the original query and our probability distribution.
We then normalise these scores and interpolate using $\lambda \in [0,1]$. In practise, we find $\lambda=0.5$ is reasonable across all datasets (see Section~\ref{sec:params} for details) .

\subsubsection{Adaptive Expansion with LEE}

We formalise adaptive expansion with LEE following our proposed expansion pipeline in Section~\ref{sec:pipeline}. Specifically, given the original query $Q$ and the current re-ranked documents, $D_{nlm}$, we produce our duet representation, $Q_{LEE}$, to retrieve the next batch of unscored, $b$-sized documents to be re-score, $D_{exp}$. Thus, as more documents are scored, and the $D_{nlm}$ set increases in size, our word and entity-based probabilistic query is updated and becomes more representative.

\section{Experimental setup}

\textbf{We release runs and hyperparameters for reproducibility:~\href{https://www.dropbox.com/s/4bmlgqyhw6k2go0/SIGIR.zip?dl=1}{\textcolor{blue}{\textit{link}}}}. Additionally, on paper acceptance we will release all code and data.
\label{sec:results}
\subsection{Data Evaluation}
\label{sec:datasets}

\subsubsection{Knowledge Base}
\label{sec:el}

Entity links provide structured connections between the queries, documents and entities. We use KILT~\cite{petroni2021kilt} to ground documents, which uses the 2019/08/01 Wikipedia snapshot containing around 5.9M entities (Wikipedia articles).  

\subsubsection{Retrieval Corpora}
We evaluate using two test collections that focus on challenging and complex information needs:

\textbf{TREC Robust04} \cite{Voorhees_TREC2004_robust} focuses on poorly performing document ranking topics.
This dataset comprises 249 topics, containing short keyword ``titles'' and longer natural-language ``descriptions'' of the information needs.  
Scaled relevance judgments are over a collection of 528k long newswire documents (TREC Disks 4 and 5).
We use 5-fold cross-validation with standard folds in previous work \cite{huston2014parameters}.

\textbf{CODEC}~\cite{mackie2022codec} focuses on the complex information needs of social science researchers  (economists, historians, and politicians).
This resource contains 42 essay-style topics and encompasses two aligned tasks: document ranking over a focused web corpus (750k long documents) and entity ranking grounded to KILT~\cite{petroni2021kilt}. 
We use the folds outlined within the online resource for 4-fold cross-validation.

\subsubsection{Indexing and Retrieval} We use Pyserini~\cite{lin2021pyserini} version 0.16.0 for indexing the corpora and datasets for terms and entities. For words, we remove stopwords and use Porter stemming. 
We store the respective entity mentions using KILT's ids as unique terms for our entity-centric document and passage indexes. 

\subsubsection{Evaluation}
We assess the system runs to a run depth of 1,000. 
We focus on recall-oriented evaluation; thus, the primary measure for this paper is Recall@1000. 
Additionally, we report MAP to understand average precision across relevant documents, and NDCG to understand the distribution of the highest-ranked documents.
We use ir-measures for all our evaluation \cite{macavaney2022streamlining}.
Lastly, we select a single baseline system for our statistical testing and use a 5\% paired-t-test significance using the scipy Python package \cite{virtanen2020scipy}.

\subsection{Entity Linking}

Previous studies show that high-recall information extraction techniques are required for successful usage in ranking tasks~\cite{kadry2017open}.
Thus, we use WAT~\cite{piccinno2014tagme} for wikification~\cite{mihalcea2007wikify} to ground both concepts and traditional named entities to Wikipedia pages. 
WAT is the successor of TagME~\cite{ferragina2011fast} and offers reasonable effectiveness that can be efficiently run over corpora-scale inputs.
Additionally, we run a state-of-the-art end-to-end entity linker ELQ~\cite{li2020efficient} over the queries, which is optimized to provide entity links for questions.
Prior work has shown enriching queries with entity-based information improves adhoc ranking performance~\cite{dalton2014entity, shehata2022early}.



\subsection{LEE Components and Hyperparameters}
\label{sec:lee-impelment}

\textbf{Neural Re-ranker (NLM)} We use T5-3b \cite{nogueira2020document}, a state-of-the-art neural re-ranking model that casts text re-ranking into a sequence-to-sequence setting.
Following the paper \cite{nogueira2020document}, we shard documents in passages of 10 sentences with a stride length of 5 and use a max-passage approach to score documents.
We use the Huggingface \texttt{castorini/monot5-3b-msmarco-10k} model.
We use the same passage shards to construct query-specific knowledge for efficiency and to align the NLM score for passage expansion methods. 

\textbf{Retrieval and Expansion} To avoid query drift, all LEE runs in the paper use the tuned BM25 system based on the input initial run~\cite{robertson1994some}.
We tune LEE hyperparameters using a grid search and cross-validation to optimise R@1000. Specifically, for our term and entity system weighting, we tune feedback passages (i.e. $fb\_docs$:  10 to 100 with a step size of 10), the number of feedback terms (i.e. $fb\_terms$:  10 to 100 with a step size of 10), the interpolation between the original terms and expansion terms (i.e. $original\_query\_weight$: 0.1 to 0.9 with a step of 0.1). For the entity component, we tune the co-occurrence weighting ($\beta$: 0.1 to 0.9 with a step of 0.1), and lastly, the hybrid weighting between word and entity ($\lambda$: 0.1 to 0.9 with a step of 0.1) and the run depth ($k_{LEE}$: 1000, 2000, 3000, and 4000). All hyperparameters are released for reproducibility: \href{https://www.dropbox.com/s/4bmlgqyhw6k2go0/ICTIR.zip?dl=1}{\textcolor{blue}{\textit{link}}}

\textbf{Adaptive Expansion} we follow the same experimental setup as \citet{sean2022adaptive} to allow a fair comparison.
Specifically, we can take an initial BM25 run ($R_0$) and use a batch size $b$ of 16 to alternate between the initial BM25 run and LEE retrieval with a total re-ranking budget of 1,000 documents. 
We use same experimental setup for the adaptive re-ranking experiments and the tuned LEE hyperparameters from initial retrieval.

\subsection{Comparison Methods}
\label{sec:comparison}

\subsubsection{First-Pass Retrieval}


\noindent \textbf{BM25}~\cite{robertson1994some}: 
Base retrieval used for expansion approaches, where we tune $k1$ (0.1 to 5.0 with a step size of 0.2) and $b$ (0.1 to 1.0 with a step size of 0.1).

\noindent \textbf{BM25 $\Rightarrow$ Relevance Model (RM3)}~\cite{abdul2004umass}: 
For BM25 with RM3 expansion, we tune $fb\_terms$ (5 to 95 with a step of 5), $fb\_docs$ (5 to 100 with a step of 5), and $original\_query\_weight$ (0.1 to 0.9 with a step of 0.1). 
This is our primary expansion baseline, separately tune RM3 expansion parameters on top of the NLM re-ranked run.

\noindent \textbf{Latent Concept Expansion (LCE)}~\cite{metzler2007latent}: This is a generalization of Relevance Modeling to include concepts as well as an IDF feature factor. We use the same tuning parameters sweeps as RM3. We compare with LCE variants that use both words and entity vocabularies. Note, the entity variants have not previously been published. 

\noindent  \textbf{SPLADE}~\cite{formal2021splade}: We use the first-passage runs provided by the author, from checkpoint: \texttt{naver/splade-cocondenser-ensembledistil}.

\noindent  \textbf{ColBERT-TCT (TCT)}~\cite{lin-etal-2021-batch}: applies knowledge distillation over the late-interaction ColBERT model~\cite{khattab2020colbert}. 
We use TCT-ColBERT-v2-HNP's model in a max-passage approach for document retrieval. 
We shard documents into passages of 10 sentences with a stride length of 5, encoding the title within each passage. 

 \noindent \textbf{ColBERT-TCT with PRF (TCT $\Rightarrow$ PRF)}~\cite{Li2021PseudoRF}:
We use the dense PRF approach by \cite{Li2021PseudoRF} and adopt their default parameters, i.e. Rocchio PRF depth is 5, $\alpha$ is 0.4 and $\beta$ 0.6. 
Furthermore, we also implement a ColBERT-TCT PRF system on top of neural re-ranking, where we take the top 5 NLM scored passages as context.

\noindent  \textbf{ENT}~\cite{shehata2022early}: 
We follow \cite{shehata2022early} recent work where queries and documents are enriched with entity context.
We re-implement a system that follows their best standalone method, ``Entities'', where we expand queries and documents with the unique names of linked entities, using ELQ for queries and WAT for documents. 
We parameter-tune BM25 in the same manner as our term-based BM25. 

\noindent  \textbf{ENT $\Rightarrow$ RM3}: 
We extend \citet{shehata2022early} to use RM3 expansion and tune parameters in the same manner as BM25 $\Rightarrow$ RM3.

\noindent  \textbf{CEQE}~\cite{naseri2021ceqe}: utilizes query-focused contextualized vectors for expansion. Specificially, we use the CEQE-MaxPool(fine-tuned) run for initial retrieval comparison and (BM25+CEDR)+CEQE-MaxPool+CEDR for query expansion post NLM re-ranking comparison.

\subsubsection{Re-Ranking}


\noindent \textbf{Entity Query Feature Expansion (EQFE)}~\cite{dalton2014entity}: 
Uses entity-based feature expansion and we include the best performing EQFE Robust04 run that is provided by the author.
 
\noindent \textbf{NLM (T5-3B)}~\cite{nogueira2020document}: 
We use a state-of-the-art sequence-to-sequence model and  follow the same setup as described in Section \ref{sec:lee-impelment}.  

\noindent \textbf{CEDR}~\cite{macavaney2019cedr}: We use the CEDR-KNRM variant with BERT-base embedding \cite{DBLP:conf/naacl/DevlinCLT19} for document ranking.

\noindent \textbf{PARADE}~\cite{li2020parade}: Uses attention to aggregate passage-level LLM signals. We use the runs from the ELECTRA-Base variant \cite{clark2020electra}.


\subsubsection{Adaptive Re-Ranking}


\noindent \textbf{GAR-BM25 $\Leftrightarrow$ NLM} \cite{sean2022adaptive}: 
We modify this adaptive passage re-ranking approach by \citet{sean2022adaptive} for document ranking.   
Specifically, we issue the re-ranked document terms as a BM25 query against the document index to identify the most similar documents.

\noindent \textbf{GAR-TCT $\Leftrightarrow$ NLM}~\cite{sean2022adaptive} 
We use ColBERT-TCT dense representations to calculate document-to-document similarity.
We take the mean of each document's passage vectors as the query vector and do a max-passage exhaustive search over the faiss index.

\noindent~\textbf{GAR-ENT $\Leftrightarrow$ NLM} 
We extend \cite{sean2022adaptive} to represent documents using the WAT document entity links.
We issue re-ranked document entities as a BM25 query against the document entity index.

\noindent~\textbf{RM3} For a fair adaptive comparison to LEE, we also use the tuned RM3 model for adaptive expansion, i.e. issue an expanded word-based query after NLM re-ranking batches.



\section{Results and Analysis}


\subsection{Research Questions}


\begin{itemize}[leftmargin=*]


\item \textit{\textbf{RQ1}: How effective is unsupervised query expansion using NLM feedback?} We compare the effectiveness of expansion models with NLM feedback, alternating the expansion models, feedback units, and vocabulary.

\item \textit{\textbf{RQ2}: How does query expansion with a second-pass NLM re-ranking compare to state-of-the-art ranking pipelines?} Here, we re-rank our LEE expansion run again and compare to entity-centric, neural re-ranking, and other PRF models with NLM feedback.  

\item \textit{\textbf{RQ3}: Does adaptive expansion provide
effectiveness gains without re-ranking more
documents?} This question focused on combining an adaptive re-ranking framework with LEE to achieve state-of-the-art effectiveness without additional NLM compute. 

\end{itemize}

\begin{table*}[h!]
\setlength{\belowcaptionskip}{-5pt}
\caption{ Query expansion varying expansion (e.g. RM3, LCE, and LEE), NLM feedback (e.g. documents ($^D$) or passages), and vocabulary (e.g. ``Entity'' or words).
Significance testing against BM25 $\Rightarrow$ RM3 $\Rightarrow$ NLM;  significantly better (``$^+$'') and worse (``$^-$''). 
}
\label{tab:doc-vs-pass}
\small
\begin{tabular}{ll|lll|lll|lll|}
\cline{3-11}
                                               &                                  & \multicolumn{3}{c|}{Robust04 - Title}                    & \multicolumn{3}{c|}{Robust04 - Description}              & \multicolumn{3}{c|}{CODEC}                           \\ \cline{3-11} 
                                               &                                  & NDCG               & MAP            & R@1000             & NDCG               & MAP            & R@1000             & NDCG           & MAP            & R@1000             \\ \hline
\multicolumn{1}{|l|}{\multirow{11}{*}{1x Re-Rank}} & BM25 $\Rightarrow$ RM3 $\Rightarrow$ NLM       & 0.634              & \textbf{0.377} & 0.777              & 0.652              & \textbf{0.406} & 0.750              & 0.644          & \textbf{0.377} & 0.816              \\ \cline{2-11} 
\multicolumn{1}{|l|}{}                         & BM25 $\Rightarrow$ NLM $\Rightarrow$ RM3-Entity$^D$ & 0.600$^-$          & 0.322$^-$      & 0.779              & 0.619$^-$          & 0.343$^-$      & 0.781$^+$          & 0.590$^-$      & 0.292$^-$      & 0.851$^+$          \\
\multicolumn{1}{|l|}{}                         & BM25 $\Rightarrow$ NLM $\Rightarrow$ RM3-Entity     & 0.612$^-$          & 0.331$^-$      & 0.776              & 0.643              & 0.364$^-$      & 0.792$^+$          & 0.645          & 0.331$^-$      & 0.854$^+$          \\

\multicolumn{1}{|l|}{}                         & BM25 $\Rightarrow$ NLM $\Rightarrow$ RM3$^D$   & 0.630              & 0.350$^-$      & 0.813$^+$          & 0.616$^-$          & 0.334$^-$      & 0.780$^+$          & 0.615          & 0.312$^-$      & 0.865$^+$          \\
\multicolumn{1}{|l|}{}                         & BM25 $\Rightarrow$ NLM $\Rightarrow$ RM3       & 0.638              & 0.353$^-$      & 0.812$^+$          & 0.625$^-$          & 0.339$^-$      & 0.797$^+$          & 0.641          & 0.335          & 0.874$^+$          \\ \cline{2-11} 
\multicolumn{1}{|l|}{}                         & BM25 $\Rightarrow$ NLM $\Rightarrow$ LCE-Entity$^D$ & 0.614$^-$          & 0.335$^-$      & 0.797              & 0.640              & 0.360$^-$      & 0.806$^+$          & 0.578$^-$      & 0.283$^-$      & 0.849              \\
\multicolumn{1}{|l|}{}                         & BM25 $\Rightarrow$ NLM $\Rightarrow$ LCE-Entity     & 0.626              & 0.343$^-$      & 0.793              & 0.659              & 0.377$^-$      & 0.810$^+$          & 0.643          & 0.325$^-$      & 0.857$^+$          \\
\multicolumn{1}{|l|}{}                         & BM25 $\Rightarrow$ NLM $\Rightarrow$ LCE$^D$   & 0.636              & 0.353$^-$      & 0.824$^+$          & 0.659              & 0.375$^-$      & 0.829$^+$          & 0.606$^-$      & 0.313$^-$      & 0.872$^+$          \\
\multicolumn{1}{|l|}{}                         & BM25 $\Rightarrow$ NLM $\Rightarrow$ LCE       & 0.647              & 0.360$^-$      & 0.825$^+$          & 0.668              & 0.377$^-$      & 0.843$^+$          & 0.632          & 0.326$^-$      & 0.877$^+$          \\ \cline{2-11} 
\multicolumn{1}{|l|}{}                         & BM25 $\Rightarrow$ NLM $\Rightarrow$ LEE$^D$        & 0.648              & 0.366          & 0.834$^+$          & 0.673$^+$          & 0.388          & 0.845$^+$          & 0.619          & 0.321$^-$      & 0.879$^+$          \\
\multicolumn{1}{|l|}{}                         & BM25 $\Rightarrow$ NLM $\Rightarrow$ LEE (Ours)     & \textbf{0.660$^+$} & 0.376          & \textbf{0.837$^+$} & \textbf{0.687$^+$} & 0.401          & \textbf{0.855$^+$} & \textbf{0.663} & 0.357          & \textbf{0.883$^+$} \\ \hline
\end{tabular}
\end{table*}

\subsection{RQ1: Expansion with NLM Feedback}
\label{sec:RQ2}

Table \ref{tab:doc-vs-pass} compares the effectiveness of expansion models on top of an NLM run, building upon the most effective sparse approaches identified previously.
Specifically, we compare a BM25 with RM3 expansion and neural re-ranking to our expansion models with NLM~\cite{nogueira2020document} feedback.
We vary the expansion models (RM3, LCE), the unit of feedback (documents and passages), and vocabulary (words and entities).
Our proposed LEE hybrid model that combines word and entity vocabularies is the last two rows in the table. 

\textbf{RQ2a: Are passages or documents more effective for neural expansion?} 
Across both datasets, we see average relative improvement of passages (rows without $^D$) to particularly improve NDCG (i.e. Robust04 titles +1.8\%, descriptions +2.4\%, and CODEC +7.2\%) and MAP (i.e. Robust04 titles +2.0\%, descriptions +3.2\%, and CODEC +10.2\%), with less relative improvement at R@1000.
This shows that passages with NLM scoring provide a more fine-grained relevance signal for our query expansion, potentially reducing noise from long documents with less relevant passages. 
We note that CODEC contains many long, domain-specific documents with over 450 non-stop words on average (roughly 50\% more than Robust04), and passages have a larger relative improvement.

\textbf{RQ2b: How does LEE's word-entity expansion compare with existing expansion approaches?}
Across all datasets, LEE has the best R@1000 and NDCG of any expansion method. Specifically, it has significantly better R@1000 compared to our NLM re-ranking baseline, with between 7.5-13.7\% relative improvement.
NDCG significantly improves on Robust04 titles and descriptions and shows relative improvements on CODEC of 2.9\% (although not significant). 
LEE is the only query expansion technique where MAP across all datasets is not significantly worse than the base neural re-ranking pipeline.
To the best of our knowledge, LEE results are the best reported first pass R@1000 results across all datasets. 
It highlights the strong recall-oriented effectiveness of word-entity hybrid models that build on neural passage re-ranking. Further, it shows the results don't hurt and sometimes help the precision; thus, additional re-ranking is not necessarily required. 

We conduct a query-by-query analysis to understand why LEE has such significant improvements in R@1000.
Focusing on Robust04 and comparing to BM25, we find that LEE helps 166 and hurts 33 title queries, compared to RM3, which helps 139 and hurts 47 queries.
These findings are even more evident in description queries, where LEE helps 181 and hurts 30 queries, compared to RM3, which helps 156 and hurts 45 queries.
This supports that LEE query expansion that leverages a combination of both words and entities is more robust than simply using words alone.

Further analysis shows that the largest relative gains are on the hardest queries (ordered based on original BM25 retrieval effectiveness).
Specifically, LEE improves R@1000 of the hardest 5\% of queries by around 0.6 compared to BM25 and 0.55 compared to BM25 with RM3 expansion.
Furthermore, substantial gains over RM3 are also observed in 5-25\% (+0.2) and 25-50\% (+0.1) buckets.
This highlights that using top-$k$ passages and joint modeling of terms and entities effectively improves the hardest queries, with minimal drop in effectiveness on the easy queries (i.e. we only slightly reduce effectiveness on 75\%-95\% band).

\textbf{RQ2c: Does entity dependencies help our query expansion model?}
Our results also show that entity-based and hybrid expansion models benefit more from passage feedback.
On average, across the datasets, such models see around 100\% greater improvement on NDCG and 150\% on MAP from using passages relative to word-based representations.
Furthermore, we find that fine-grained passage signals are important for leveraging entity information, especially when using dependencies to infer relationships between entities.
We find that including the entity co-occurrences improves effectiveness versus simply modelling entities based on unigrams; they provide consistent improvements across the datasets increasing MAP by 3.3\% on average, NDCG 0.9\% and R@1000 0.4\%, with no system being negatively affected. 
However, we find that entity co-occurrence at a document level is less effective, with MAP reducing on average by 0.2\%, with small gains in NDCG and R@100 of 0.3\%, and Robust04 systems being negatively affected. 

Overall, LEE improves recall-oriented effectiveness by leveraging NLM re-ranked passages to infer a strong hydrid word and entity relevance model.
We show that it is more effective than either RM3 and LCE on words and entities individually, and that entity dependencies help with passage feedback.
Our unsupervised LEE queries are significantly better across all datasets on R@1000, either substantially improving or not significantly hurting MAP and NDCG.
These are the best-reported R@1000 results across these datasets and highlight the effectiveness of hybrid word-entity expansion models in combination with NLM passage ranking. 


\begin{table*}[h!]
\setlength{\belowcaptionskip}{-5pt}
\caption{
Expansion with NLM feedback and second-pass re-ranking;
``+'' significant improvement over BM25 $\Rightarrow$ RM3 $\Rightarrow$ NLM.}
\label{tab:single-shot-results}
\small
\begin{tabular}{ll|lll|lll|lll|}
\cline{3-11}
                                              &                                                & \multicolumn{3}{c|}{Robust04 - Title}                        & \multicolumn{3}{c|}{Robust04 - Description}                  & \multicolumn{3}{c|}{CODEC}                               \\ \cline{3-11} 
                                              &                                                & NDCG               & MAP                & R@1000             & NDCG               & MAP                & R@1000             & NDCG               & MAP            & R@1000             \\ \hline
\multicolumn{1}{|l|}{\multirow{9}{*}{1x Re-Rank}} & SPLADE~\cite{formal2021splade}  $\Rightarrow$ NLM         & 0.539	
       & 0.309      & 0.597      & 0.590             & 0.357             & 0.617              & \multicolumn{1}{l}{-} & \multicolumn{1}{l}{-} & \multicolumn{1}{l|}{-} \\
\multicolumn{1}{|l|}{}  & EQFE~\cite{dalton2014entity}                                           & 0.601              & 0.328              & 0.806              & -                  & -                  & -                  & -                  & -              & -                  \\
\multicolumn{1}{|l|}{}                        & BM25 $\Rightarrow$ CEQE~\cite{naseri2021ceqe} $\Rightarrow$ NLM      & 0.626          & 0.373          & 0.764          & \multicolumn{1}{l}{-} & \multicolumn{1}{l}{-} & \multicolumn{1}{l|}{-} & \multicolumn{1}{l}{-} & \multicolumn{1}{l}{-} & \multicolumn{1}{l|}{-} \\

\multicolumn{1}{|l|}{}                        & BM25 $\Rightarrow$ RM3 $\Rightarrow$ 
 CEDR~\cite{macavaney2019cedr}                                          & 0.632              & 0.370              & 0.776              & 0.645              & 0.400              & 0.758              & -                  & -              & -                  \\
\multicolumn{1}{|l|}{}                        & BM25 $\Rightarrow$ RM3 $\Rightarrow$ 
 PARADE~\cite{li2020parade}                                         & 0.642              & 0.380              & 0.776              & 0.650              & 0.408              & 0.758              & -                  & -              & -                  \\

\multicolumn{1}{|l|}{}                        & ENT $\Rightarrow$ RM3 $\Rightarrow$ NLM & 0.615          & 0.366      & 0.745      & 0.658                 & 0.407        & 0.759         & 0.490                 & 0.373                 & 0.833         \\
\multicolumn{1}{|l|}{}                        & TCT $\Rightarrow$ PRF $\Rightarrow$ NLM        & 0.584          & 0.345      & 0.681      & 0.572    & 0.364             & 0.619              & 0.606                 & 0.351             & 0.754              \\

\multicolumn{1}{|l|}{}                        & BM25 $\Rightarrow$ RM3 $\Rightarrow$ NLM                     & 0.634              & 0.377              & 0.777              & 0.652              & 0.406              & 0.750              & 0.644              & 0.377          & 0.816              \\
\multicolumn{1}{|l|}{}                        & BM25 $\Rightarrow$ NLM $\Rightarrow$ LEE (Ours)                   & 0.660$^+$          & 0.376              & \textbf{0.837$^+$} & 0.687$^+$          & 0.401              & \textbf{0.855$^+$} & 0.663              & 0.357          & \textbf{0.883$^+$} \\ \hline
\multicolumn{1}{|l|}{\multirow{4}{*}{2x Re-Rank}}  & CEDR $\Rightarrow$ CEQE~\cite{naseri2021ceqe}   $\Rightarrow$ NLM     & 0.644              & 0.384              & 0.787              & -                  & -                  & -                  & -                  & -              & -                  \\
\multicolumn{1}{|l|}{}                       & TCT $\Rightarrow$ NLM $\Rightarrow$ PRF $\Rightarrow$ NLM    & 0.592              & 0.349              & 0.697              & 0.630              & 0.390              & 0.702              & 0.636              & 0.369          & 0.808              \\
\multicolumn{1}{|l|}{}                        & BM25 $\Rightarrow$ NLM $\Rightarrow$ RM3 $\Rightarrow$ NLM        & 0.656$^+$          & 0.390$^+$          & 0.813$^+$          & 0.674$^+$          & 0.416$^+$          & 0.780$^+$          & 0.659              & 0.379          & 0.865$^+$          \\
\multicolumn{1}{|l|}{}                        & BM25 $\Rightarrow$ NLM $\Rightarrow$ LEE $\Rightarrow$ NLM (Ours) & \textbf{0.667$^+$} & \textbf{0.393$^+$} & \textbf{0.837$^+$} & \textbf{0.715$^+$} & \textbf{0.438$^+$} & \textbf{0.855$^+$} & \textbf{0.664$^+$} & \textbf{0.380} & \textbf{0.883$^+$} \\ \hline
\end{tabular}
\end{table*}

\subsection{RQ2: Second-Pass Re-Ranking Effectiveness}
\label{sec:RQ3}

This research question addresses how the LEE expansion model with passage feedback compares to sparse and dense systems with an additional round of neural re-ranking.
Specifically, Table \ref{tab:single-shot-results} shows LEE (with a second re-ranking phase~\cite{nogueira2020document}) compared to current state-of-the-art neural and traditional models on the target datasets.
This includes CEDR~\cite{macavaney2019cedr} and PARADE~\cite{li2020parade}, which are a strong systems fine-tuned on Robust04.  

We also compare our system to comparable neural pseudo-relevance feedback techniques that leverage multiple rounds of re-ranking.
We include the CEQE, which used CEDR's initial re-ranked run, and we also implement a ColBERT-TCT-PRF and RM3 expansion runs on top of a NLM (T5-3b) run, with all systems NLM re-ranked for a fair comparison.
We conduct significance testing against BM25 with RM3 expansion NLM re-ranking system.

These results highlight how gains in effectiveness can be achieved with NLM feedback across standard sparse and dense PRF retrieval models. Specifically, we see relative recall improvements of 5\% with RM3 expansions and 8\% with ColBERT-TCT-PRF using NLM feedback.   
Moreover, a second pass neural re-ranker over our LEE initial retrieval run further improves NDCG and MAP.
This leads to NDCG and R@1000 being significantly improved compared to the state-of-the-art baseline, with MAP significantly better on Robust04 titles and descriptions.
Sadly, MORES \cite{gao2022long} only reports NDCG@20; however, given our recall improvements, LEE with NLM would almost certainly have higher NDCG and MAP.  
Additionally, LEE with re-ranking achieves the best MAP and NDCG scores when compared to state-of-the-art prior methods and varying NLM expansion methods (CEQE, ColBERT-PRF, and RM3).

\begin{table*}[b!]
\setlength{\belowcaptionskip}{-5pt}
\caption{
Adaptive re-ranking effectiveness (``$\Leftrightarrow$''), with significance testing (``+'') against BM25 $\Rightarrow$ RM3 $\Rightarrow$ NLM.}
\label{tab:iterative-results}
\small
\begin{tabular}{ll|lll|lll|lll|}
\cline{3-11}
                                                                                                    &                                                & \multicolumn{3}{c|}{Robust04 - Title}                        & \multicolumn{3}{c|}{Robust04 - Description}                  & \multicolumn{3}{c|}{CODEC}                               \\ \cline{3-11} 
                                                                                                    &                                                & NDCG               & MAP                & R@1000             & NDCG               & MAP                & R@1000             & NDCG               & MAP            & R@1000             \\ \hline
\multicolumn{1}{|l|}{2x Re-Rank}                                                                        & BM25 $\Rightarrow$ NLM $\Rightarrow$ LEE $\Rightarrow$ NLM (Ours) & 0.667$^+$          & \textbf{0.393$^+$} & 0.837$^+$          & \textbf{0.715$^+$} & \textbf{0.438$^+$} & \textbf{0.855$^+$} & 0.664$^+$          & 0.380          & 0.883$^+$          \\ \hline
\multicolumn{1}{|l|}{\multirow{2}{*}{1x Re-Rank}}                                                                        & BM25 $\Rightarrow$ RM3 $\Rightarrow$ NLM                     & 0.634              & 0.377              & 0.777              & 0.652              & 0.406              & 0.750              & 0.644              & 0.377          & 0.816              \\
\multicolumn{1}{|l|}{}                        & BM25 $\Rightarrow$ NLM $\Rightarrow$ LEE (Ours)                   & 0.660$^+$          & 0.376              & 0.837$^+$ & 0.687$^+$          & 0.401              & \textbf{0.855$^+$} & 0.663              & 0.357          & 0.883$^+$ \\ \hline

\multicolumn{1}{|l|}{\multirow{5}{*}{\begin{tabular}[c]{@{}l@{}}1x Re-Rank\\ (Adaptive)\end{tabular}}} & BM25 $\Rightarrow$ GAR-BM25 $\Leftrightarrow$ NLM                 & 0.629              & 0.372              & 0.768              & 0.652              & 0.402              & 0.747              & 0.634              & 0.362          & 0.797              \\
\multicolumn{1}{|l|}{}                                                                              & BM25 $\Rightarrow$ GAR-ColBERT $\Leftrightarrow$ NLM              & 0.630              & 0.374              & 0.769              & 0.649              & 0.402              & 0.739              & 0.645              & 0.368          & 0.822              \\
\multicolumn{1}{|l|}{}                                                                              & BM25 $\Rightarrow$ GAR-ENT $\Leftrightarrow$ NLM                  & 0.637              & 0.377              & 0.781              & 0.661              & 0.408              & 0.758              & 0.644              & 0.366          & 0.821              \\
\multicolumn{1}{|l|}{}                                                                              & BM25 $\Rightarrow$ RM3 $\Leftrightarrow$ NLM                      & 0.655$^+$          & 0.387$^+$          & 0.813$^+$          & 0.675$^+$          & 0.418$^+$          & 0.783$^+$          & 0.653              & 0.373          & 0.847              \\
\multicolumn{1}{|l|}{}                                                                              & BM25 $\Rightarrow$ LEE $\Leftrightarrow$ NLM  (Ours)                    & \textbf{0.668$^+$} & 0.392$^+$          & \textbf{0.838$^+$} & 0.704$^+$          & 0.435$^+$          & 0.834$^+$          & \textbf{0.669$^+$} & \textbf{0.382} & \textbf{0.887$^+$} \\ \hline
\end{tabular}
\end{table*}

Although neural re-ranking improves MAP and NDCG results, it is worth highlighting how competitive the LEE unsupervised expansion method (without re-ranking) is when compared to prior work.
Specifically, neural re-ranking only increases NDCG between 0.002-0.028 and MAP 0.018-0.037 across the datasets.
This improvement from NLM re-ranking is roughly five times less than neural re-ranking over a BM25 with RM3 expansion run.
For example, using the unsupervised LEE query on Robust04 titles, we achieve NDCG@10 of 0.561, which is higher than reported SPLADE~\cite{formal2021splade} results 0.485 (a comparable unsupervised method), better than T5-3b~\cite{nogueira2020document} 0.545, and approaching fine-tune PARADE~\cite{li2020parade} 0.591.

Query analysis shows LEE outperforms on entity-centric queries, where the topical focus is a specific concept or named entity, where dense models typically struggle~\cite{sciavolino2021simple}.
For example, in Robust04 query ``World Court'', LEE can use the entity mentions to model the probability of the central entity, [International\_Court\_of\_Justice], and increases R@1000 by at least 0.5 compared to sparse and dense baselines.
On further analysis, LEE's effectiveness within entity-centric information needs is not surprising when we analyse the implicit entity ranking from LEE using CODEC's entity judgements.
We can find that LEE is very effective in the top ranks, achieving NDCG@3 of 0.767 and NDCG@10 of 0.554 (much higher than all dataset baselines).

Overall, these results highlight the potential of NLM feedback to improve recall across PRF models and improve state-of-the-art effectiveness.
We show how a second pass re-ranking can further improve LEE's precision and that our hybrid word-entity modelling helps entity-centric queries.
Nonetheless, doing a second-pass NLM scoring increases the computational re-ranking expense, which we seek to address in the following research question.

\subsection{RQ3: Adaptive Expansion}
\label{sec:RQ4}

In this research question, we explore combining LEE with adaptive expansion to improve effectiveness without a second pass re-ranking.
Table \ref{tab:iterative-results} shows adaptive LEE expansion against LEE with two NLM passes, the adaptive ``GAR'' systems \cite{sean2022adaptive}, and an adaptive RM3 expansion system for comparison.
We conduct significance testing against BM25 with RM3 expansion with NLM re-ranking.


\textbf{RQ3a: Is adaptive expansion with LEE the most effective adaptive re-ranking method?}
We find that GAR-based methods that use words (GAR-BM25), entities (GAR-ENT), and dense representation (GAR-ColBERT) are not significantly better than a standard NLM re-ranking pipeline.
Given GAR's strong effectiveness on passage ranking, these results suggest document ranking over long documents requires relevance modelling across multiple documents.
In fact, even adaptive expansion with RM3 is consistently more effective than all GAR systems, being significantly better on Robust04 over the NLM re-ranking pipeline and better, although not significantly, on CODEC.


Moreover, these results support LEE with adaptive expansion as the most effective adaptive method across all datasets and measures.
The significance testing aligns with two re-ranking phases, i.e. being significantly better on Robust04 across all measures and CODEC on R@1000 and NDCG. 
In fact, LEE with adaptive expansion is nominally better than two re-ranking passes on CODEC across all measures and Robust04 titles on NDCG and R@1000.
Analysis shows that iteratively sampling different documents after each batch can lead to a better language model.

Although, R@1000 reduces on Robust04 descriptions from 0.853 to 0.834 when comparing adaptive re-ranking and two NLM passes. On inspection, this reduction in recall effectiveness is driven by alternating batches between BM25 and LEE following \citet{sean2022adaptive} algorithm.
Because LEE is much more relatively effective than BM25, we see a reduction in overall effectiveness. 
Nonetheless, adaptive re-ranking does improve NDCG by 2.5\% and MAP by 6.6\% over any state-of-the-art methods on description queries, and future work could vary how we sample from these different pools.

\textbf{RQ3b: What are the computational benefits of adaptive expansion?}
For simplicity, we measure computational expense by the number of documents that require NLM re-ranking, which should be a strong proxy across implementations and hardware.
Thus, the computations benefits of adaptive expansion are due to the document set differences between the initial run (i.e. BM25) and the LEE with NLM feedback (i.e. BM25 $\Rightarrow$ NLM $\Rightarrow$ LEE). For example, on Robust04 titles, two passes of NLM results in 1,503 unique documents being scored per query, compared to only 1,000 for adaptive re-ranking (i.e. saves 33\% scoring cost). We find similar trends in Robust04 descriptions (637 fewer documents to re-score) and CODEC (525 fewer documents to re-score).   
Overall, we show that LEE, as part of an adaptive expansion framework, can significantly improve recall-oriented effectiveness without requiring a second pass of neural re-ranking.



\subsection{Discussion of Parameters}
\label{sec:params}

\begin{table}[b!]
\setlength{\belowcaptionskip}{-5pt}
\caption{
Tuned LEE model vs zero-shot LEE model (CODEC parameters); ``+'' and ``-'' are significance testing against tuned.}
\label{tab:zero-shot}
\small
\begin{tabular}{lllllll}
\cline{2-7}
\multicolumn{1}{l|}{}                            & \multicolumn{3}{c|}{Robust04 - Titles}      & \multicolumn{3}{c|}{Robust04 - Descriptions} \\ \cline{2-7} 
\multicolumn{1}{l|}{\textit{1x Re-Rank}} & NDCG  & MAP   & \multicolumn{1}{l|}{R@1000} & NDCG   & MAP   & \multicolumn{1}{l|}{R@1000} \\ \hline
\multicolumn{1}{|l|}{Tuned}                      & 0.660  & 0.376 & \multicolumn{1}{l|}{0.837}  & 0.687  & 0.401 & \multicolumn{1}{l|}{0.855}  \\
\multicolumn{1}{|l|}{Zero-Shot}                  & 0.660  & 0.374 & \multicolumn{1}{l|}{0.836}  & 0.688  & 0.401 & \multicolumn{1}{l|}{0.846}  \\ \hline
\textit{2x Re-Rank}                      &       &       &                             &        &       &                             \\ \hline
\multicolumn{1}{|l|}{Tuned}                      & 0.667 & 0.393 & \multicolumn{1}{l|}{0.837}  & 0.715  & 0.438 & \multicolumn{1}{l|}{0.855}  \\
\multicolumn{1}{|l|}{Zero-Shot}                  & 0.667 & 0.394 & \multicolumn{1}{l|}{0.836}  & 0.710  & 0.435 & \multicolumn{1}{l|}{0.846}  \\ \hline
\textit{Adaptive}                      &       &       &                             &        &       & \multicolumn{1}{l|}{}       \\ \hline
\multicolumn{1}{|l|}{Tuned}                  & 0.668 & 0.392 & \multicolumn{1}{l|}{0.838}  & 0.704  & 0.435 & \multicolumn{1}{l|}{0.834}   \\
\multicolumn{1}{|l|}{Zero-Shot}                  & 0.668  & 0.393  & \multicolumn{1}{l|}{0.837}  & 0.701  & 0.432 & \multicolumn{1}{l|}{0.829}  \\ \hline
\end{tabular}
\end{table}

As outlined in Section~\ref{sec:lee-impelment}, we tune our LEE model following the official cross-validation setup outlined for target datasets. However, here we analyse: (1) how effective our method is zero-shot and (2) the impact of $\lambda$, i.e., the relative weighting of words and entities.    

To assess LEE expansion in a zero-shot scenario, we use LEE parameters tuned on the CODEC dataset zero-shot on Robust04 titles and descriptions (the exact parameters can be found the released run metadata).
Table~\ref{tab:zero-shot} shows the ``Tuned'' LEE expansion model against the ``Zero-shot'' parameters for our unsupervided LEE query, two rounds of NLM re-ranking, and adaptive expansion.
We observed no significant differences between the tuned and zero-shot LEE run under these different scenarios, and in some cases zero-shot is the same or marginally better on Robust04 titles.
Therefore, this highlights that our proposed method of using NLM passage feedback and combining words and entities with dependencies is robust to its parameter across datasets.

\begin{figure}[h!]
    \centering
    \setlength{\abovecaptionskip}{-1pt}
    \setlength{\belowcaptionskip}{-5pt}
    \includegraphics[scale=0.33]{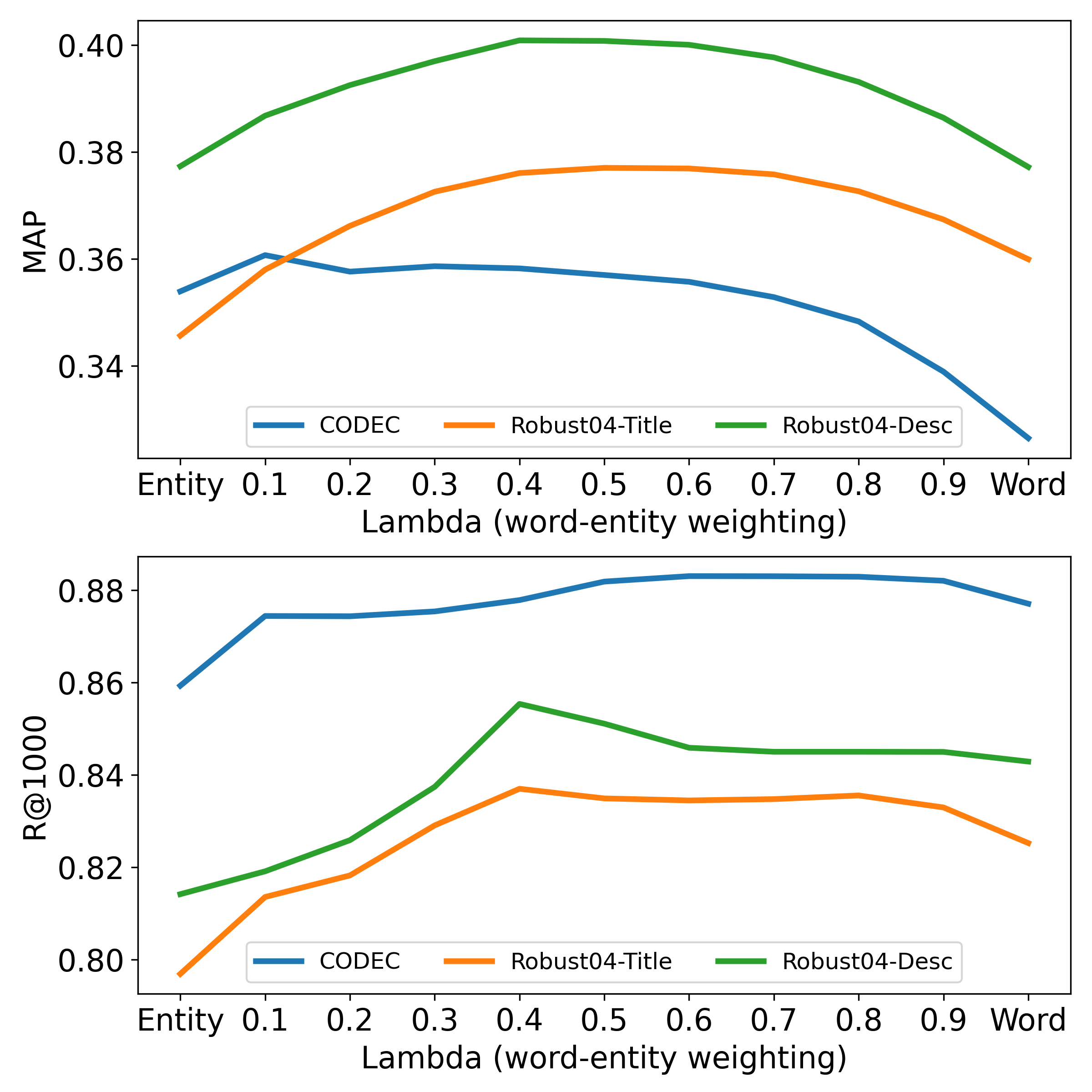}
    \caption{
    Lambda (i.e. relative word-to-entity weighting) impact for LEE expansion on CODEC and Robust04 datasets.   
    }
    \label{img:lambda}
\end{figure}

Figure \ref{img:lambda} shows the impact of lambda (i.e. relative word-to-entity weighting) on the effectiveness of LEE unsupervised query across our target datasets.
Specifically, we see that for R@1000 and MAP, the best weighting is a combination of words and entities.
For Robust04 datasets, MAP maximizes around 0.5, which weights word and entity expansions equally. However, for CODEC, precision is maximized around 0.1, favouring weighting entities and showing their precision benefits on domain-specific essay questions.
On the other hand, Robust04 shows optimal recall with a relatively even combination of words and entities. However, unlike MAP, R@1000 for CODEC is maximized through a high weighting of words.
Overall, this should show the precision-recall tradeoffs for different datasets and confirms that both words and entities are required for robust effectiveness.


\section{Conclusion}
\label{sec:conclusion}

We show that the LEE word-entity expansion using fine-grain passage feedback from NLM re-ranking significantly improves R@1000, with between 8-14\% improvement over RM3 expansion.
Specifically, the joint modelling of words and entities at a passage level improves relevance modelling, including incorporating entity dependencies. 
Our method is robust in terms of query-level hurts vs helps, improves recall of the hardest queries by 0.6, and can use parameters across datasets without significantly harming effectiveness.
Additionally, we show that our implicit entity ranking is highly effective within the top ranks and helps improve entity-centric queries.
Lastly, we demonstrate that LEE with adaptive expansion can avoid two NLM passes and achieve state-of-the-art effectiveness without additional document re-ranking (saving 35\% of the re-ranking cost).
We believe adaptive expansion can lead to new dynamic expansion models to improve both effectiveness and efficiency.

\bibliographystyle{ACM-Reference-Format}
\balance
\bibliography{foo}

\end{document}